\documentclass{pasj00}

\begin{document}
\SetRunningHead{Ishiyama {\it et al.}}{Environmental effect on the subhalo abundance}
\Received{}
\Accepted{}

\title{
Environmental effect on the subhalo abundance\\
--- a solution to the missing dwarf  problem.
}


\author{Tomoaki \textsc{Ishiyama}}
\affil{National Astronomical Observatory, Mitaka, Tokyo 181-8588, Japan\\
and Department of General System Studies, College of Arts and Sciences,\\
University of Tokyo, Tokyo 153-8902, Japan}
\email{ishiyama@cfca.jp}

\author{Toshiyuki \textsc{Fukushige}}
\affil{K\&F Computing Research Co., Chofu, Tokyo 192-0026, Japan}
\email{fukushig@kfcr.jp}

\and

\author{Junichiro \textsc{Makino}}
\affil{National Astronomical Observatory, Mitaka, Tokyo 181-8588, Japan}
\email{makino@cfca.jp}


%


\maketitle

\begin{abstract}

Recent high-resolution simulations of the formation of dark-matter
halos have shown that the distribution of subhalos is scale-free, in
the sense that if scaled by the velocity dispersion of the parent
halo, the subhalo velocity distribution function of galaxy-sized and
cluster-sized halos are identical. For cluster-sized halos, simulation
results agreed well with observations. Simulations, however, predicted
far too many subhalos for galaxy-sized halos. Our galaxy has several
tens of known dwarf galaxies. On the other hands, simulated
dark-matter halos contain thousands of subhalos.  We have performed
simulation of a single large volume and measured the abundance of
subhalos in all massive halos. We found that the variation of the
subhalo abundance is very large, and those with largest number of
subhalos correspond to simulated halos in previous studies.  The
subhalo abundance depends strongly on the local density of the
background. Halos in high-density regions contain large number of
subhalos. Our galaxy is in the low-density region. For our simulated
halos in low-density regions, the number of subhalos is within a
factor of four to that of our galaxy.  We argue that the ``missing
dwarf problem'' is not a real problem but caused by the biased
selection of the initial conditions in previous studies, which were
not appropriate for field galaxies.

\end{abstract}

\section{Introduction}

Klypin et al. (1999) and Moore et al.(1999a) analyzed the structure of
dark-matter halos formed in their high-resolution cosmological
$N$-body simulations and found that dark matter halos of the mass
comparable to the Local group contain far too many subhalos compared
to known dwarf galaxies in the Local group. This ``missing dwarf
problem'' was confirmed by many simulations based on the concordance
cosmological model, and has been regarded as one of its most serious
problems.  A number of ``solutions'' have been proposed, including ``warm'' dark matter
(Kamionkowski, Liddle 2000), self-interacting dark matter(Spergel,
Steinhardt 2000), suppression of star formation by early reionization
(Susa, Umemura 2004), self-regulation of star formation in small halos
(Stoehr et al. 2002; Kravtsov et al. 2004), but
none is widely accepted as a clear-cut solution.

In this paper, we study the environmental effect on the subhalo
abundance. In almost all previous studies of substructures in the
dark-matter halos, simulations are done following the
``re-simulation'' prescription, in which one first simulates a fairly
large volume (for galaxy-sized halos typically a 50-100Mpc cube) with
low mass resolution and identifies the candidate regions to perform
high-resolution simulations. Then, one makes the same initial model, but
with higher mass resolution for the candidate regions, and analyze the
result. Clearly, this is not the way to obtain an unbiased sample of
dark-matter halos.

In order to see if the bias has any effect, we simulated a
relatively small region (a 21.4 Mpc box) with high mass resolution
(mass of particles = $3\times 10^{6}M_{\odot}$) and high spacial
resolution (softening length= 1.8kpc). Number of particles used is
$512^3$. Simulation was performed with TreePM code (Yoshikawa,
Fukushige 2005) on a 12-node GRAPE-6A cluster (Sugimoto, et al. 1990;
Fukushige et al. 2005). Since the simulated region is still relatively
small, we might be affected from the bias due to the fact we neglected
the contributions of fluctuations with the wavelength longer than the
vbox size, but we are free from the halo-selection bias, since we
analyzed all halos with rotation velocity larger than 200 km/s.  There
are 21 such halos and we give them ID numbers in order of their
masses.

\begin{figure*}[t]
\centering
\includegraphics[width=16cm]{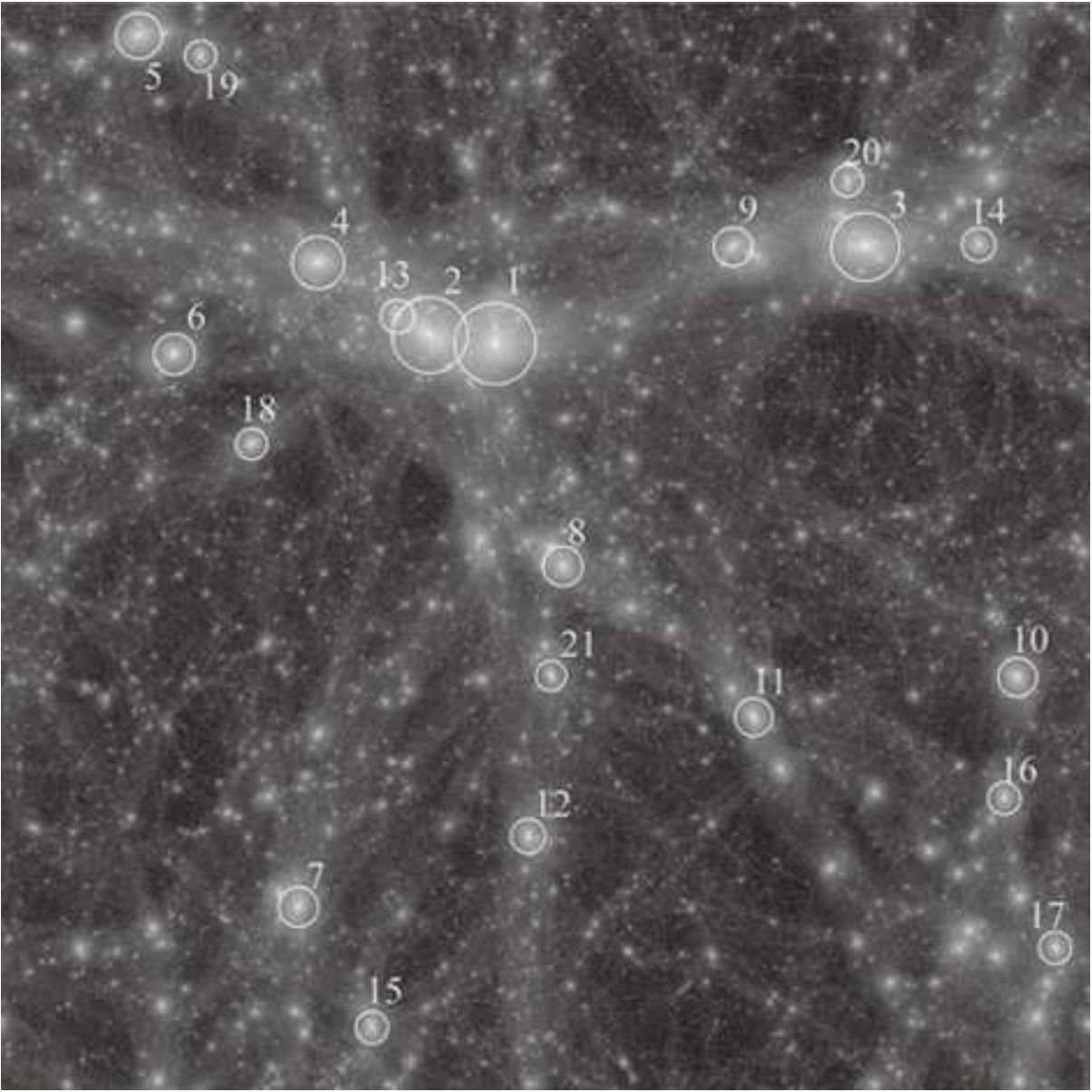}
\caption{Snapshot of the simulation box whose size is 21.4Mpc at $z=0$. 
The circles shows the virial radii of the 21 selected halos. 
The numbers near circles indicate the halo IDs. }
\label{fig1}
\end{figure*}

The structure of this paper is as follows.  In section 2, we describe
the method of our $N$-body simulation.  In section 3, we present the
results of simulation.  Section 4 is for conclusion.

\section{Method}

We performed an LCDM ($\Omega_0=0.3$, $\lambda_0=0.7$, $h=0.7$,
$\sigma_8=1.0$) cosmological simulations.  Here, $\Omega_0$ is the
density parameter, $\lambda_0$ is the dimensionless cosmological
constant, $H_0 = 100 h \, {\rm km/s}\cdot{\rm Mpc}^{-1}$ at the
present epoch, and $\sigma_8$ is the top-hat filtered mass variance at
$8 h^{-1}$ Mpc. The initial particle distribution was generated using
GRAFIC1 code (Bertschinger 1995).

We followed the evolution of $512^3$ particles with masses of $3\times
10^6 M_{\odot}$ in a comoving 21.4 Mpc cube, using a parallel TreePM
code (Yoshikawa, Fukushige 2005) on a GRAPE-6A cluster (Fukushige,
Makino, Kawai 2005).  We set the grid size for the PM part as $256^3$,
and the opening parameter for the tree part as $\theta=0.5$.

We integrated the system in the comoving coordinates with a leap-flog
integrator with shared timestep.  The step size is adaptive and
determined according to $\min(2.0\sqrt{\varepsilon/|\vec{a}_i|},
2.0\varepsilon/|\vec{v}_i|)$, We integrated the system from $z=84$ to 0
and the total number of timesteps was $4930$.  The (Plummer)
gravitational softening we used is constant in the comoving coordinate
upto $z=10$, and is constant (1.8 kpc) in the physical coordinate from
$z=10$ to $z=0$.

\begin{figure*}[t]
\centering
\includegraphics[width=16cm]{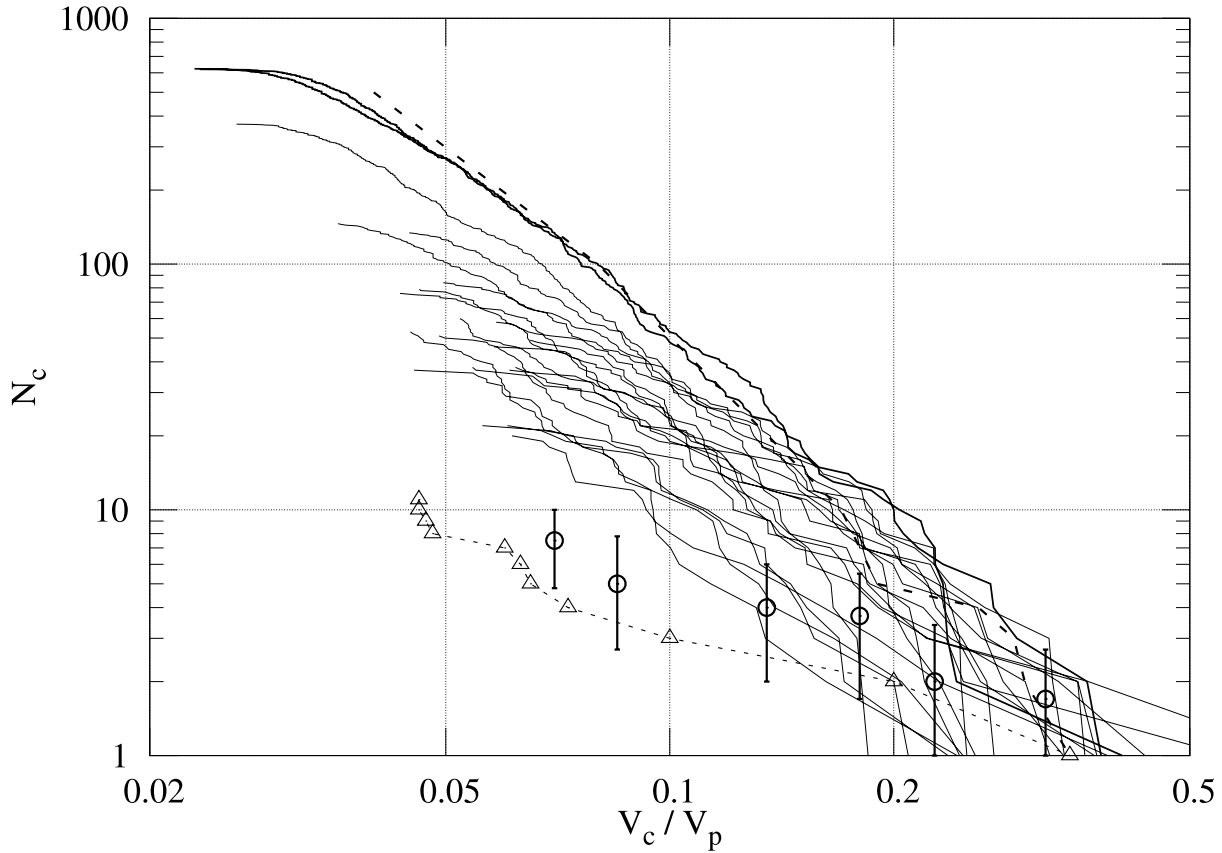}
\caption{Cumulative numbers of subhalos as a function of their maximum
circular velocities $v_{\rm c}$ normalized by those of the parent halos
$v_{\rm p}$ for all 21 selected halos. Two upper bold curves are for
those of group-sized halos G1 and G2.  The thick dashed
curve is the result of Moore et al. (1999) for a galaxy-sized halo.
The thin dashed curve with open triangles denotes the dwarf galaxies
in our galaxy (Mateo 1998). 
Open circles with error bar show the dwarf galaxies in the local group (D'Onghia et al. 2007).
}
\label{fig2}
\end{figure*}

We used a 12-node GRAPE-6A cluster at University of Tokyo. The parallel
cluster consists of 12 host computers (Pentium 4/2.8GHz, i865) each of
which has one GRAPE-6A board.  The simulation presented below needs
$\sim 280$ seconds per timestep, and thus one run ($4930$ timesteps)
is completed in $380$ hours (wallclock time).

In Table \ref{tab1}, we summarize the mass $M$, the radius $r_{\rm
v}$, the maximum rotation velocity, $v_{\rm p}$, the total number of
particles, $N$, and the number of particles in the $N_{>0.1}$-th
subhalo, $N_{{\rm sub},0.1}$, of the 21 selected halos.  We also
summarize the three dimensional position ($X/L,Y/L,Z/L$) of the halos in
the simulation box, where we set the origin of coordinate at the
bottom-left corner of figure \ref{fig1}.

\begin{figure*}[t]
\centering
\includegraphics[width=16cm]{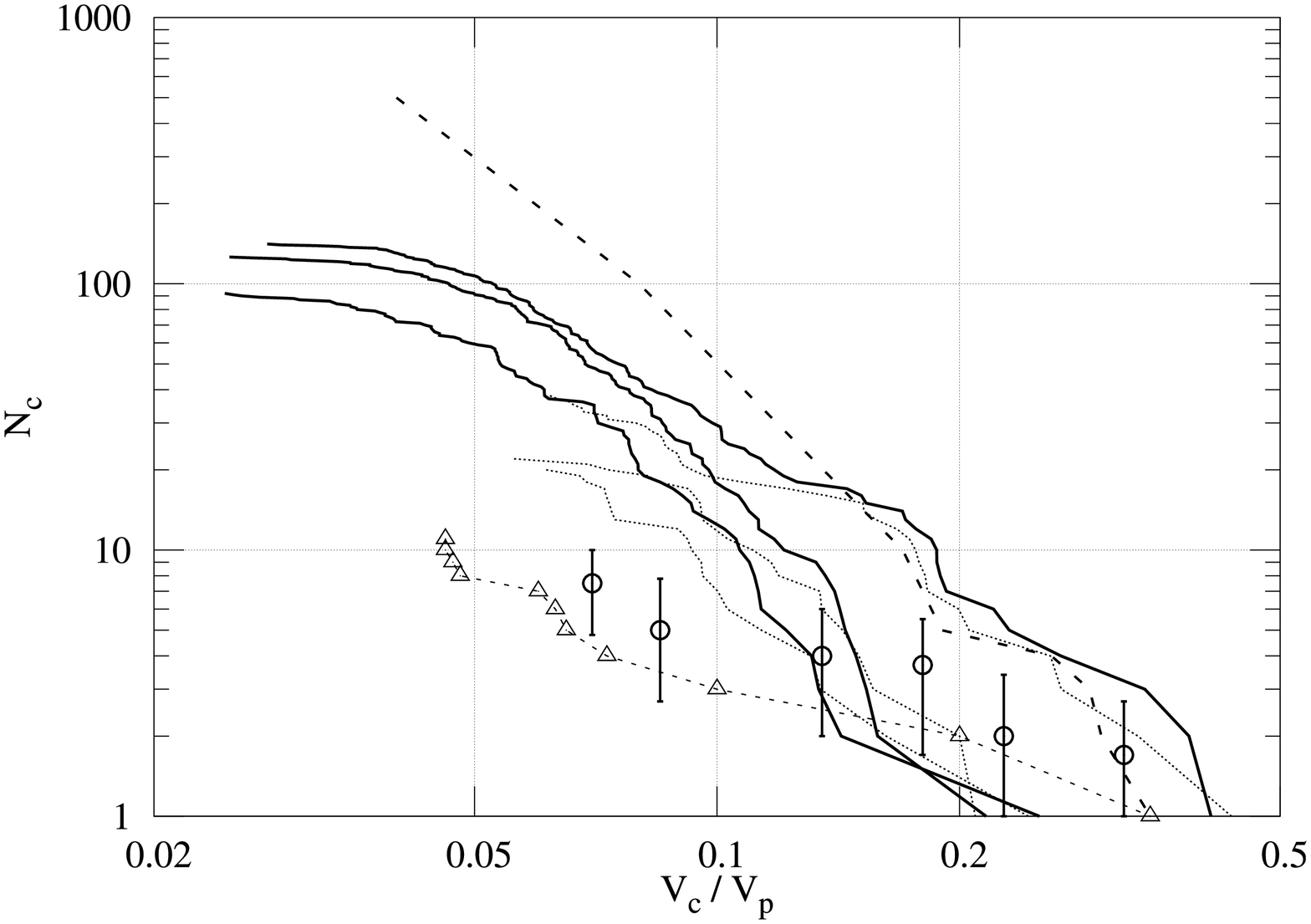}
\caption{Same as figure 1 for two subhalo-poor halos G16, G20 and a
  subhalo-rich one G18.  
Thick curves are the results of the higher resolution simulation.
Thin dotted curves are the results of the original simulation.
From top down, the results of G18, G20, G16 are plotted.
The thick dashed curve is the result of Moore et al. (1999) for a galaxy-sized halo.
The thin dashed curve with open triangles denotes the dwarf galaxies
in our galaxy (Mateo 1998). 
Open circles with error bar show the dwarf galaxies in the local group (D'Onghia et al. 2007).
}
\label{fig3}
\end{figure*}

In order to investigate the effect of mass and spatial resolution on 
the subhalo abundance, we performed one simulation in which several
subhalos are replaced with higher-resolution ones.
We selected
two subhalo-poor halos (G16 and G20) and one subhalo-rich one (G18).  We
picked particles within 5$r_{\rm v}$ from the center of these halos at
z=0, and traced these particles back to the initial condition.  We
replaced these particles with higher-resolution particles
($3.7\times 10^5 M_{\odot}$), and re-ran the simulation.  The total
number of timesteps was 9869.  The gravitational softening we used was
constant in the comoving coordinates for $z>10$, and was constant (0.89
kpc) in the physical coordinates for $z<10$.  We labeled
these halos HG16, HG18, HG20.  In the bottom of Table \ref{tab1}, we
summarize the properties of these halos.

\section{Result}

\begin{figure*}[t]
\centering
\includegraphics[width=5.5cm,angle=270]{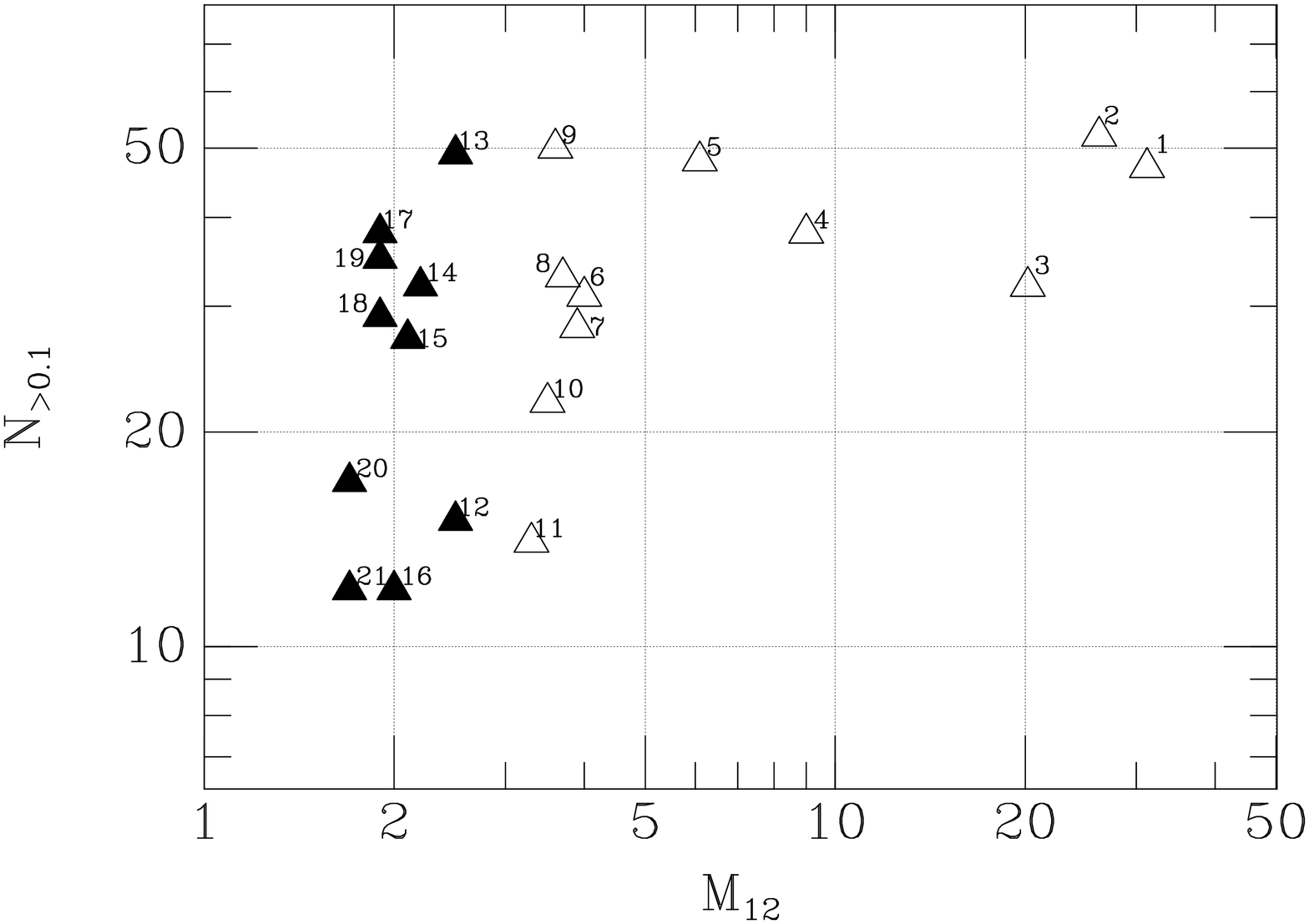}
\includegraphics[width=5.5cm,angle=270]{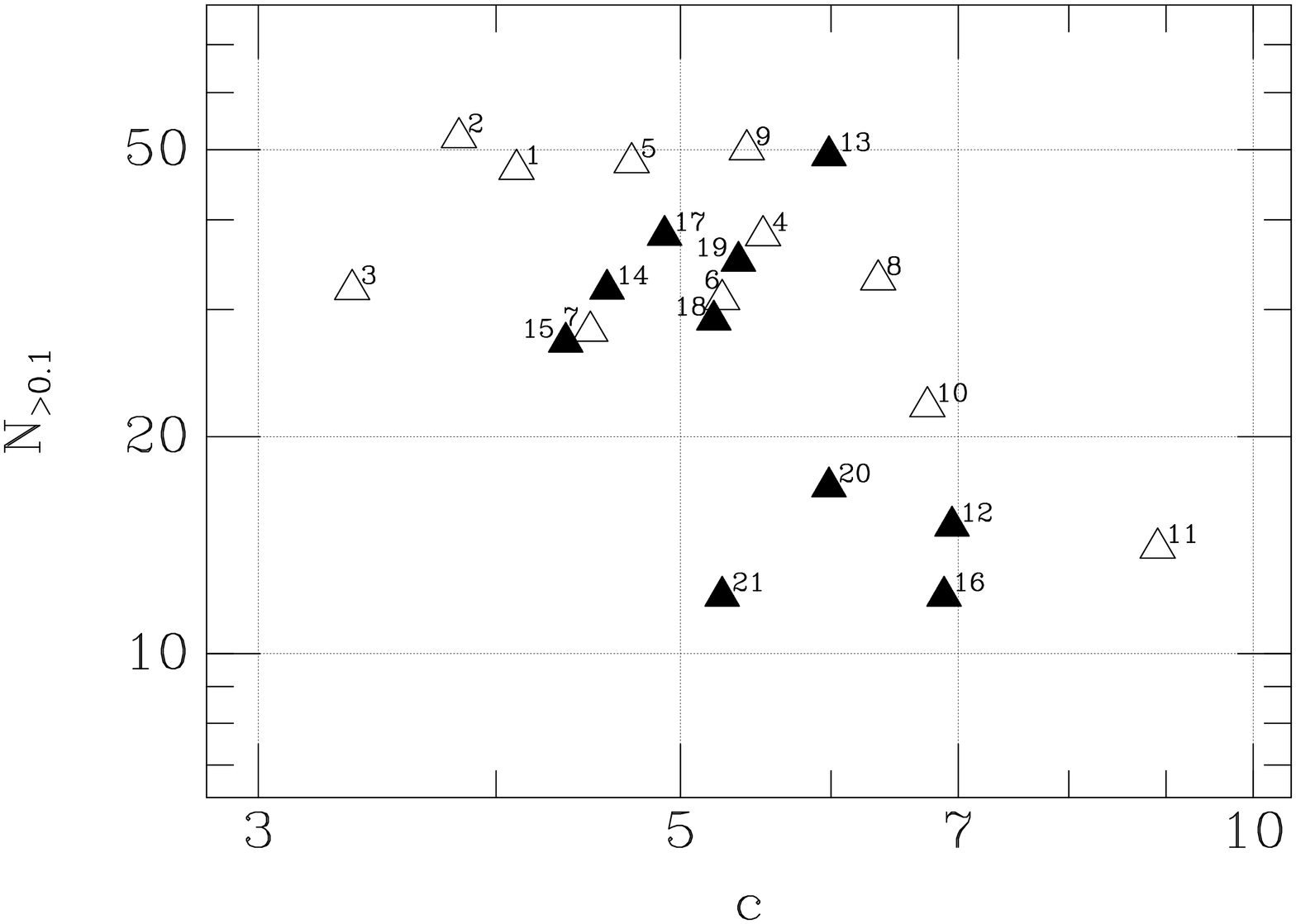}

\includegraphics[width=5.5cm,angle=270]{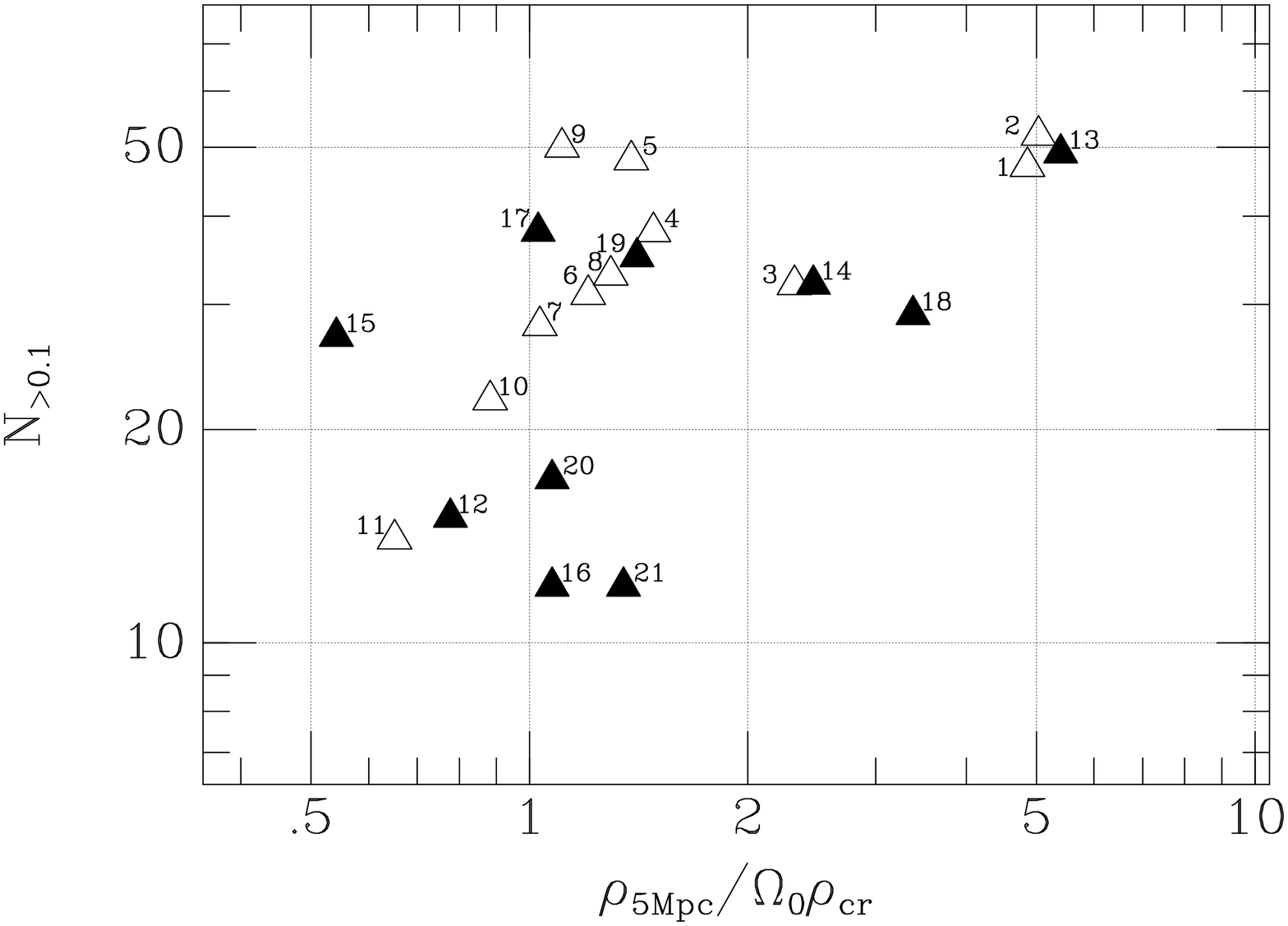}
\includegraphics[width=5.5cm,angle=270]{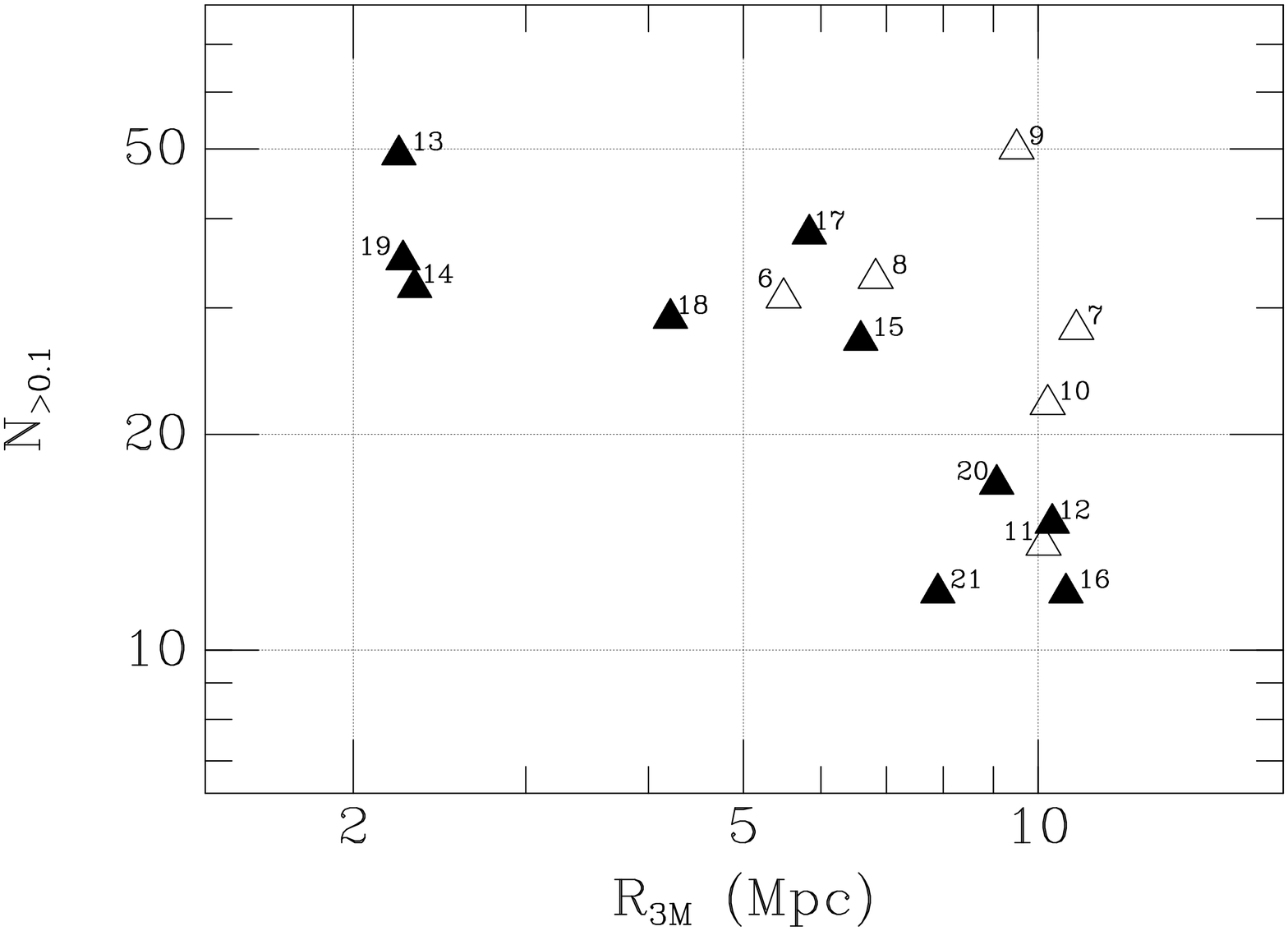}

\caption{Dependence of subhalo abundance $N_{>0.1}$ on (a) mass $M_{12}$ in unit
of $10^{12}M_{\odot}$ (upper-left), (b) concentration parameter
$c=r_0/r_{\rm v}$ (upper-right), (c) local density within 5Mpc
(lower-left), and (d) distance to influential larger-sized halos,
$R_{3M}$, of parent halos (lower-right).  The numbers near symbols indicate the halo IDs.
The black triangle symbols mean the halos of $M<3\times
10^{12}M_{\odot}$. }
\label{fig4}
\end{figure*}

\begin{figure*}[t]
\centering
\includegraphics[width=8cm]{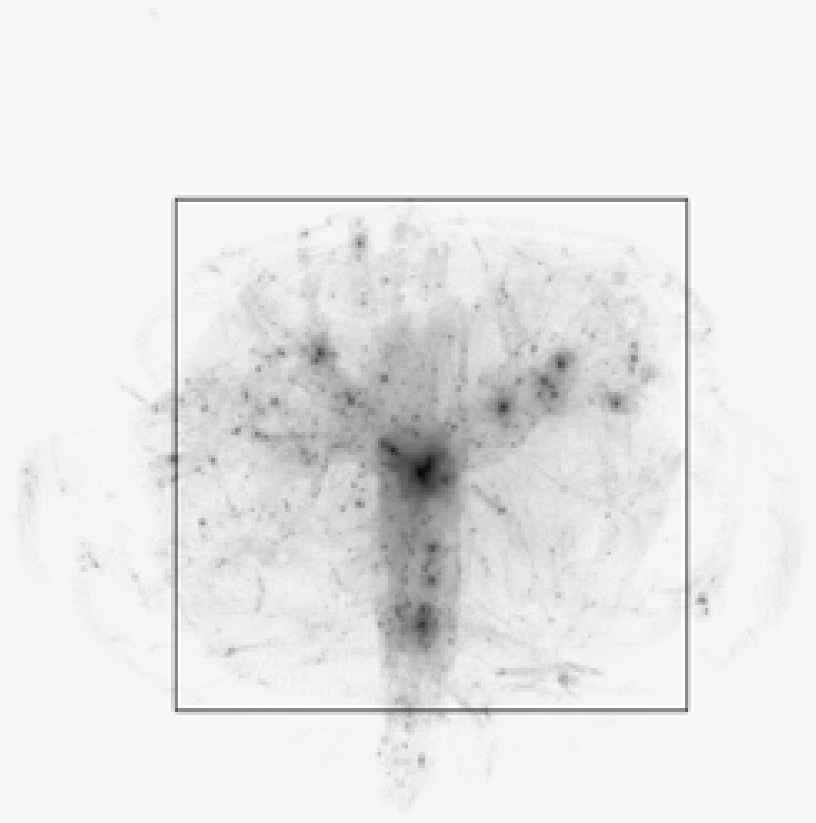}
\includegraphics[width=8cm]{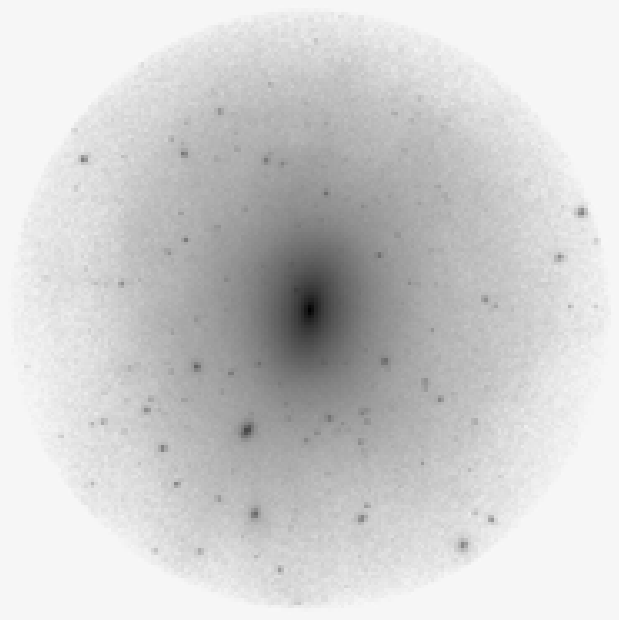}

\includegraphics[width=8cm]{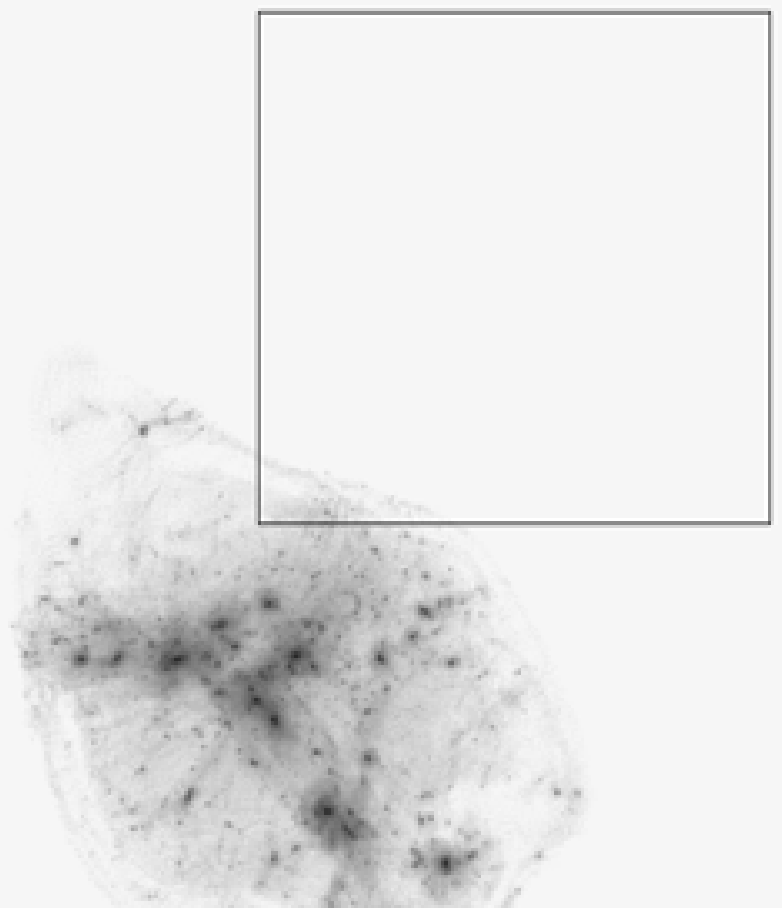}
\includegraphics[width=8cm]{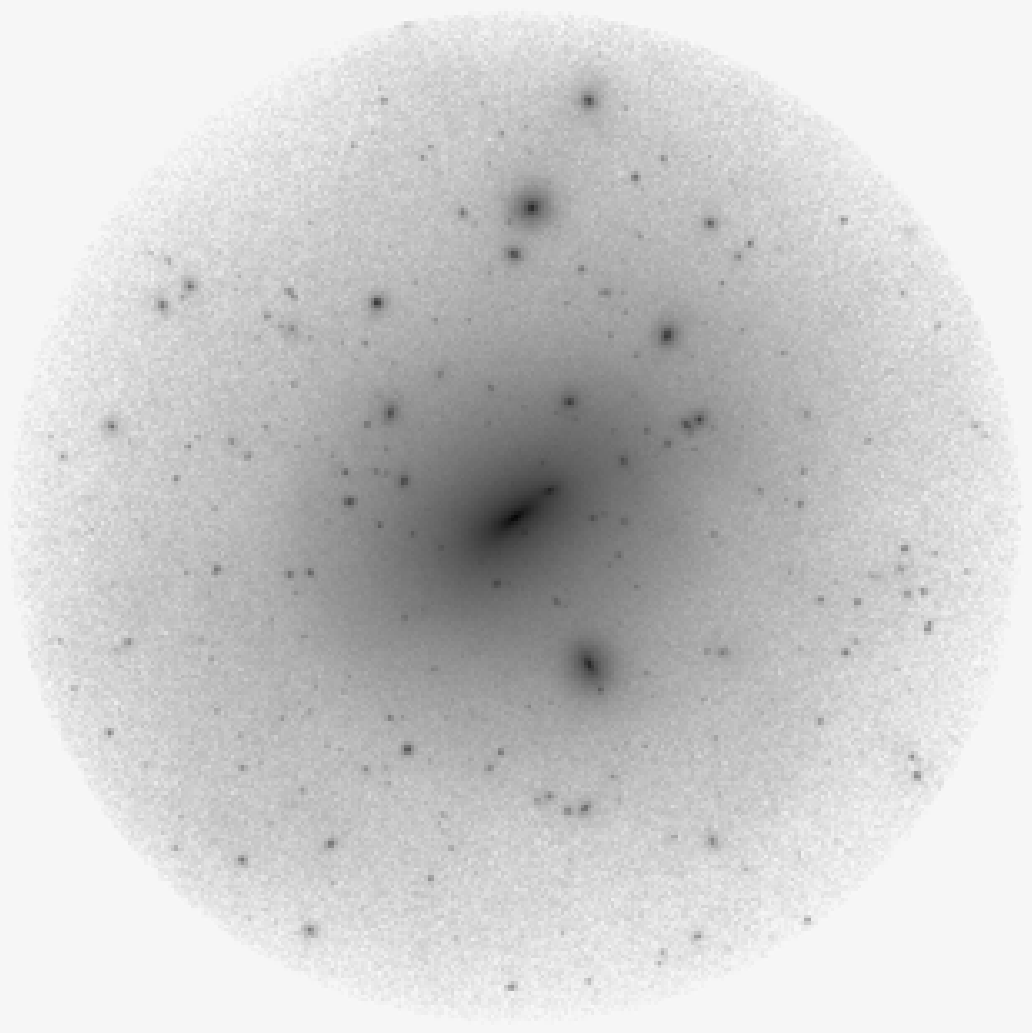}

\caption{Snapshots of a subhalo-poor halo G16 and a subhalo-rich
 one G18 (upper to lower) at z=3 (left) and z=0 (right).
The width is 2Mpc (left) and 0.8Mpc (right).
The boxes on the left snapshots show the regions of the right snapshots.}
\label{fig5}
\end{figure*}

Figure \ref{fig1} shows the particle distribution in the simulation
box at $z=0$.  The length of the side $L$ is 21.4 Mpc. We unbiasedly
selected 21 (parent) halos for which the maximum rotation velocity
exceeds 200km/s.  The rotation velocity is defined as $v_{\rm cir}(r)
= \sqrt{GM(r)/r}$.

Figure \ref{fig2} shows the cumulative distribution of
subhalos as the function of circular velocities
for the selected 21 halos together. We identify all local potential
minima within the virial radius $r_{\rm v}$ as subhalos. In Figure
\ref{fig2}, we plot cumulative numbers of subhalos as a function of
their maxima of circular velocity, $v_{\rm c}$, scaled by those of the
parent halos $v_{\rm p}$. Two upper bold curves indicate those of
group-sized (more than $10^{13}M_{\odot}$) halos, G1 and G2.  Other
thin solid curves are for other 19 less massive halos. We can see that
the number of subhalos in these halos shows very large variation.  The
thick dashed curve is the result of Moore et al. (1999a) for a
galaxy-sized halo. Moore et al.'s halo has the similar number of subhalos
as that of the most subhalo-rich halos in our simulation. Thin curve with open
triangles denotes the dwarf galaxies in our galaxy (Mateo 1998). The
distribution of dwarf galaxies in our galaxy is not too different from
that of the most subhalo-poor halos in our simulation.
Open circles with error bar show the dwarf galaxies in the local group 
(D'Onghia et al. (2007), who gave somewhat larger number of dwarfs than Mateo (1998) did).
We can see that the difference between observation of local group and 
least subhalo-poor halos in our simulation is pretty small.

Figure \ref{fig3} shows the velocity distribution function of subhalos
for three selected halos (G16,G18 and G20) together.  Thick curves are
the results of the higher resolution simulation.  Thin dotted curves
are the results of the original simulation.  From top down, the
results of G18, G20, G16 are plotted.  The number of subhalos is
larger for the higher resolution simulation, in particular for small
subhalos.  The number of subhalos with $v_{\rm c}/v_{\rm p}>0.1$ in
high-resolution runs is
about 1.6 times larger than that in the original simulation, for all
three subhalos. Thus, large variation in the number of subhalo is also
visible in high-resolution run, and it is not a numerical artifact. On
the other hand, the original simulation gives systematically low
number of subhalos at least for these halos with small total mass. This means that
we must correct the number of subhalos in original simulation.  In
order to make a correction to the result of the original simulation using the higher
resolution simulation, we apply the following correction formula
\begin{eqnarray}
 N_{>0.1, {\rm corrected}} =& C  N_{>0.1, {\rm original}},\\
 C =& \cases{ 1.6   &$ V_p < 200{\rm km/s}$\cr
   1-0.6 \log_{2}(V_p/400) & $200< V_p < 400$\cr
   1                & $ V_p > 400$\cr
 },
\end{eqnarray}
where $N_{>0.1}$ is the number of subhalos with circular velocity more
than 10\% of that of the parent halo. We use this corrected number as
the measure of the subhalo abundance. 
We used this measure
because it is known to be safe against small-$N$ effects. Number counts for
subhalos containing less than 200 particles is not reliable (Kase et
al. 2007).  Even for G21, the subhalos with $v_{\rm c}/v_{\rm p}>0.1$
contain more than 200 particles. The systematic difference we found is
probably due to the rather large softening used in the original simulation.

In order to understand the origin of this large variation in the
subhalo abundance, we investigated the dependence of the abundance on
various quantities. Figure \ref{fig4} shows the result. 
The subhalo abundance clearly depends on the mass of the parent halo,
as seen in figure \ref{fig4}a . However, if we consider only
galaxy-sized halos (halo mass less than $5\times 10^{12}M_{\odot}$),
there remains very large scatter in the abundance. Massive halos (mass
$M> 5\times 10^{12}M_{\odot}$) are all abundant in subhalos.

Figure \ref{fig4}b shows the dependence on the concentration parameter
$c=r_0/r_v$, where $r_0$ is the break radius for the fit to Moore et
al.'s profile (Moore et al. 1999b) and $r_v$ is again the virial radius. More
centrally-concentrated halos have fewer subhalos. This dependence
might mean that the concentrated halos, which are dynamically more
evolved, have fewer subhalos.

Figure \ref{fig4}c shows the dependence on the local mass density
averaged over $5 {\rm Mpc}$. Again, there is clear correlation, and if
the parent halo is in a low density region, it contains less subhalos,
though some of low-mass halos in low-density region (G15, G17 and G19)
have relatively many halos.

Figure \ref{fig4}d shows the dependence on the distance to the nearest
halo with mass more than three times the mass of the parent halo. This
is another way to measure the effect of the environment. The dependence
is similar to that on the local density, but we can see that halos G17
and G19, which are in low-density regions but relatively rich in
subhalos, have nearby massive halos.  From figures \ref{fig4}c and
\ref{fig4}d, we can conclude that galaxy-sized halos have small number
of subhalos if they are formed in low-density region, with no nearby
massive halos.

This dependence on the local environment is somewhat
counter-intuitive. No matter what the external environment is, halos
should be formed bottom-up, and the merging history, which might have
some effect on the subhalo abundance, should not depend much on the
local density {\it outside} the halo, as far as the mass and velocity
dispersion of the halos are similar. As we can see from table 1,
velocity dispersion depends strongly on the halo mass and therefore
dependence on the local density is weak.

Figure \ref{fig5} shows
one subhalo-poor halo (G16) and one subhalo-rich one (G18)
at $z=0$ and $z=3$. For $z=3$, only particles which is in the halo's
virial radius at $z=0$ are plotted. From panels for $z=0$, we can
clearly see the large difference in the subhalo abundance. From panels
for $z=3$, we can see that they have very different
shapes. Subhalo-poor halos are much more centrally concentrated than
subhalo-rich ones. Thus, even though they initially contain large number
of substructures, many of them are disrupted or lost most of mass due
to the tidal field of the parent halo. On the other hand,
subhalo-rich halos have many substructures which are distant from the
high-density central region of the parent halo, and many of them
survived to $z=0$, simply because they remain far away from 
the high-density region of the parent halo.

The difference in the shape at $z=3$ is the direct consequence of the
difference in the local density. The regions which will become
galaxy-sized halo have similar average density at early time, whether
or not they are in the high-density region or low-density
region. However, this means that halos in the low-density region must
have more power in the small-scale density fluctuation which directly
corresponds to the halo mass, since they do not have the contribution
of large-scale fluctuations. Thus, they are more centrally
concentrated.  As we have seen above, the concentration shows as
anti-correlation with the subhalo abundance.

Another possible effect is the difference in the external tidal
field. From figure \ref{fig5}, we can see that halos G16 did not
move much from $z=3$ to $z=0$, while G18 traveled a rather large
distance. This difference is because of the difference in the external
tidal field. The external tidal field has the effect of changing the
orbits of subhalos, generally adding more angular momenta. Thus,
subhalos in the halos in high-density regions have orbits with larger
pericenter distances. 

In the case of cluster-sized or even group-sized halos, this effect of
the local environment must be relatively weak, because the density
fluctuations with mass scale larger than the cluster mass do not have
much power. Thus, this strong environmental effect is unique to the
galaxy-sized halos. The reason why this effect is overlooked in
previous studies is, in our opinion, the prejudice that the
gravitational structure formation is scale free and dark halos should
behave in the same way irrespective of their masses or environments.

\section{Conclusion}

We have performed the simulation of a single large volume and measured the
abundance of subhalos in all massive halos. We found that the
variation of the subhalo abundance is very large.  The subhalo
abundance depends strongly on the local density of the
background. Halos in high-density regions contain large number of
subhalos. Our galaxy is in the low-density region. For our simulated
halos in low-density regions, the number of subhalos is within a
factor of four to those of our galaxy or local group.

We conclude that the ``missing dwarf problem'' was to a large extent
caused by the biases introduced in the standard practice used to
prepare initial conditions for cosmological simulations, and if we
construct unbiased samples, galaxy-sized halos in low-density field
region contain subhalos whose number is not inconsistent with the
number of dwarf galaxies in the Local group.

\bigskip
We are grateful to Kohji Yoshikawa for providing his parallel TreePM code.
We thank Hiroyuki Kase and Keigo Nitadori 
for their technical advices and helpful discussions.
Numerical computations were in part carried out on the PC cluster at
Center for Computational Astrophysics (CfCA),
National Astronomical Observatory of Japan.
This research is partially supported by 
the Special Coordination Fund for Promoting Science and Technology
(GRAPE-DR project), Ministry of Education, Culture, Sports, Science and
Technology, Japan.

\begin{center}
\begin{table*}[h]
\caption{Selected Halos\label{tab1}}
\begin{tabular}{ccccrrccc}
\hline
\hline
Halo ID & $M$ & $r_{\rm v}$ & $V_{\rm p}$ 
& $N\quad$ & $N_{{\rm sub},0.1}$ 
& $X/L$ & $Y/L$ & $Z/L$ \\
& ($10^{12}M_{\odot}$) & (kpc) & (km/s) \\
\hline
G1 & 31.2 & 798 & 472 & 10466465 & 2241 & 0.454 & 0.687 & 0.637 \\
G2 & 26.2 & 752 & 421 & 8780560 &  1969 & 0.394 & 0.694 & 0.701 \\
G3 & 20.0 & 688 & 434 & 6700220 &  1625 & 0.793 & 0.776 & 0.475 \\
G4 & 9.0 & 528 & 334 & 3025883 &    993 & 0.291 & 0.761 & 0.960 \\
G5 & 6.1 & 465 & 284 & 2063080 &    527 & 0.126 & 0.969 & 0.363 \\
G6 & 4.0 & 404 & 258 & 1355302 &    423 & 0.159 & 0.677 & 0.597 \\
G7 & 3.9 & 398 & 250 & 1299438 &    373 & 0.273 & 0.169 & 0.556 \\
G8 & 3.7 & 393 & 251 & 1251621 &    404 & 0.516 & 0.482 & 0.905 \\
G9 & 3.6 & 389 & 247 & 1214417 &    343 & 0.672 & 0.775 & 0.037 \\
G10& 3.5 & 386 & 260 & 1180493 &   417 & 0.933 & 0.380 & 0.704 \\
G11 & 3.3 & 379 & 278 & 1115028 &   496 & 0.691 & 0.343 & 0.312 \\
G12 & 2.5 & 343 & 244 & 830204 &    390 & 0.484 & 0.234 & 0.815 \\
G13 & 2.5 & 343 & 219 & 828903 &    242 & 0.364 & 0.712 & 0.799 \\
G14 & 2.2 & 329 & 211 & 737135 &    230 & 0.898 & 0.778 & 0.446 \\
G15 & 2.1 & 324 & 205 & 701198 &    224 & 0.340 & 0.058 & 0.897 \\
G16 & 2.0 & 317 & 227 & 655852 &    323 & 0.921 & 0.269 & 0.705 \\
G17 & 1.9 & 317 & 202 & 653996 &    202 & 0.968 & 0.131 & 0.514 \\
G18 & 1.9 & 316 & 202 & 650187 &    208 & 0.230 & 0.594 & 0.658 \\
G19 & 1.9 & 313 & 200 & 626605 &    217 & 0.183 & 0.951 & 0.449 \\
G20 & 1.7 & 305 & 200 & 584875 &    214 & 0.777 & 0.836 & 0.894 \\
G21 & 1.7 & 301 & 200 & 559250 &    221 & 0.505 & 0.381 & 0.861 \\
HG16 & 1.9 & 315 & 228 & 5189215 &    1715 & 0.921 & 0.269 & 0.706 \\
HG18 & 1.9 & 314 & 203 & 5140043 &    1308 & 0.228 & 0.592 & 0.658 \\
HG20 & 1.7 & 303 & 200 & 4666533 &    1357 & 0.778 & 0.836 & 0.895 \\
\hline
\end{tabular}
\end{table*}
\end{center}

\end{document}